\documentclass[conference]{IEEEtran}
\ifCLASSINFOpdf
   \usepackage[pdftex]{graphicx}
\else
   \usepackage[dvips]{graphicx}
\fi
\usepackage{fixltx2e}
\usepackage{url}

\usepackage{color}
\usepackage{cite}

\begin{document}
%
\title{When Money Learns to Fly: Towards Sensing as a Service Applications Using Bitcoin}

\author{\IEEEauthorblockN{Kay Noyen\IEEEauthorrefmark{1}, Dirk Volland\IEEEauthorrefmark{2}, Dominic W\"orner\IEEEauthorrefmark{1} and Elgar Fleisch\IEEEauthorrefmark{1}\IEEEauthorrefmark{2}} \IEEEauthorblockA{\IEEEauthorrefmark{1}Department of Management, Technology and Economics\\
ETH Zurich, Switzerland}
 \IEEEauthorblockA{\IEEEauthorrefmark{2}Institute of Technology Management\\
 University of St. Gallen, Switzerland
 }
 }


%


\maketitle

\begin{abstract}
Sensing-as-a-Service (S\textsuperscript{2}aaS) is an emerging Internet of Things (IOT) business model pattern. To be technically feasible and to effectively allow for broad adoption, S\textsuperscript{2}aaS implementations have to overcome manifold systemic hurdles, specifically regarding payment and sensor identification. In an effort to overcome these hurdles, we propose Bitcoin as  protocol for S\textsuperscript{2}aaS networks. To lay the groundwork and start the conversation about disruptive changes that Bitcoin technology could bring to S\textsuperscript{2}aaS concepts and IOT in general, we identify and discuss the core characteristics that could drive those changes. We present a conceptual example and describe the basic process of exchanging data for cash using Bitcoin.

\end{abstract}


%
\IEEEpeerreviewmaketitle

\section{Introduction}
After investigating product, process and service innovation, scholars and industry experts have recently been shifting their focus on the innovation of business models, i.e. on how companies comprehensively reinvent the way they create and capture value for their customers \cite{osterwalder2010business, zott2011business, pohle2006ibm, gassmann2013st,timmers1998business}. In a majority of game changing business model innovation in the last 15 years, information technology played a key role \cite{gassmann2013st}. In that period, many new so called business model patterns where created in which information technology was a mandatory precondition. The first wave of the Internet, for example, created patterns such as \emph{Freemium}, \emph{E-Commerce}, \emph{Leverage Customer Data}, \emph{Long Tail}, and \emph{Digitalization}, the Internet 2.0 then led to patterns such as \emph{Crowdsourcing}, \emph{Crowdfunding}, and \emph{User Design}. 

Given that each Internet wave so far created new business model patterns one can assume that the same holds true for the next wave of Internet-based innovation, the \emph{Internet of Things} (IOT). The IOT is expected to consist of billions of sensor nodes bridging the gap between the physical and digital world. Hence, \emph{Sensing-as-a-Service} (S\textsuperscript{2}aaS) is a promising candidate for an IOT-enabled business model pattern \cite{weinberger,perera2014sensing, mizouni2013mobile, sheng2013sensing}.  

In contrast to a physical good, a digital service attached to it by connectivity and sensing capabilities can easily be delivered not just to one addressee, but to many customers at the same time. In addition, these customers can be distributed across the globe. As a consequence, the one who generates the data is not necessarily the one who profits from the data. Thus physical goods that help to create data initiate multi-sided markets for sensor data in which one or more customer groups (the market’s buying side) subscribe to and pay for data that is provided by one more data creator (selling side). For instance, car manufacturers could be interested in a constant flow of road condition data generated by all cars on the street, city planners in the way how bicycles move through town, parking authorities and car drivers in empty parking lots, companies that produce weather forecasts in the data of millions of privately owned weather stations, and home security systems in the data generated by heating and climate control systems.

The technical challenges when building such a S\textsuperscript{2}aaS business model pattern are manifold: For instance, sensors and its owners have to be uniquely identified and authenticated, the values delivered by the sensors should be traceable and have to be secured against manipulation, and a low cost micropayment system has to enable the seller side to receive monetary gratification from the buyer side — because not everybody will be willing to share his sensor data for free. But even more than technical challenges, economic and business aspects are central to the success of S\textsuperscript{2}aaS \cite{bohli2009initial} given the inherent chicken-and-egg problem that is typical for patterns based on network effects. In this paper, we present the key characteristics of Bitcoin -- a peer-to-peer payment system and a digital currency -- and discuss how these characteristics could contribute to enable and establish  S\textsuperscript{2}aaS business model patterns.

The remainder of this paper is structured as follows. First, we briefly introduce Bitcoin and its current direction of development and research. Second, we present the basic concept of using Bitcoin in the context of S\textsuperscript{2}aaS, further elaborating on how to scale our approach. Finally, we discuss the S\textsuperscript{2}aaS relevant core characteristics of Bitcoin, conclude our work and give directions for future research.

\section{Bitcoin Primer}
Since its invention in the year 2008 \cite{nakamoto2008bitcoin}, Bitcoin has grown to a global phenomenon allowing for frictionless, borderless transfer of cash endorsed by an enthusiastic community of early adopters, entrepreneurs, and speculators. 
Bitcoin is a peer-to-peer payment system that uses cryptography to allow payments to be sent directly from one party to another without going through a central authority like a financial institution. The term Bitcoin refers to both the technology and payment system that was introduced as open source software in 2009, and to the currency itself. Unlike traditional currencies like the Euro or US Dollar, but also unlike other digital currencies like Linden Dollars or Amazon Coins, and alternative payment services including eWallets like Paypal or money transfer systems like MoneyGram, Bitcoin does not rely on trusting a central authority that controls money supply, money distribution, or transaction verification \cite{EuropeanCentralBank}. Bitcoin solely relies on an eco-system designed and defined by the Bitcoin protocol and its open source implementations. A collectively implemented public ledger called the block chain keeps record of all transactions, and therefore allows to appropriate ownership of bitcoins at all points in time \cite{nakamoto2008bitcoin}. Digital signatures based on public- and private key pairs are used to determine and prove the ownership of units of Bitcoin \cite{salomaa1996public}. Bitcoin eliminates payment fraud, which is a highly significant and expensive issue for payment receivers (merchants) in traditional payment networks. Fraudulent credit card payments (e.g. initiated using a stolen credit card) can be reversed after the payment occurred. All losses are usually covered by the receiver \cite{schwartz1985credit}. If a transaction from one Bitcoin address to another has been carried out, it is impossible for anyone else than the receiver to reverse the transaction. Bitcoin further uses a scripting system that provides flexibility to change transactions parameters of what is needed to be able to spend transferred bitcoins \cite{smartcontr}.

Since the seminal paper from Satoshi Nakamoto \cite{nakamoto2008bitcoin}, Bitcoin has received a high level of attention from the press. Despite its current instantiation still being criticized for its poor parameters, its core has the potential for a proliferating decentralized currency \cite{barber2012bitter}. Indeed, the eco-system around Bitcoin has been growing at rapid pace \cite{Meiklejohn2013}. Popular representatives for this eco-system are BitPay \cite{bitpay.com} that enables merchants to accept bitcoins or Blockchain.info \cite{blockchain.info}, a Bitcoin wallet and block chain explorer service. Also a vast number of Bitcoin clones and derivatives like Litecoin \cite{litecoin.org} or Peercoin \cite{peercoin.net} have emerged, to name only two of them. The concept of Bitcoin has also gained popularity in academic research. A Google Scholar query for the term bitcoin indicates exponential growth of related literature, with 20 results before the year 2010, 140 results for 2011, 509 results for 2012, and 1020 results for 2013. Recent research has also vested the concept of Bitcoin, and its method for achieving consensus, beyond purposes of pure payment transactions. These include usage of Bitcoin and its block chain or parts of it for other purposes, but also the use of Bitcoin concepts to build their own separate systems.

Bitmessage \cite{warren2012bitmessage} is a trustless decentralized peer-to-peer communication system that enables encrypted communication. Applications like Colored Coins \cite{coloredcoins.org}, Mastercoin \cite{mastercoin.org}, or Counterparty \cite{counterparty.co} build on top of the bitcoin protocol and add new layers that may be used to implement commodity certificates, smart property, or instruments such as stocks or bonds \cite{Buterin2013, Rosenfeld2012}. Namecoin \cite{namecoin.info} is based on Bitcoin and allows to create globally unified namespaces that can be used to e.g. decentralize the concept of domain name systems \cite{Namecoin}. While Namecoin provides the mechanism for globally unified namespaces for all sorts of applications, OneName \cite{onename.io} uses the Namecoin block chain and defines a format for usernames and profiles. BitDeposit uses Bitcoin and proposes an economic measure to deter attacks and service abuses of cloud computing applications \cite{szefer2013bitdeposit}. Others like Ethereum \cite{ethereum.org}, referring to themselves as cryptocurrency 2.0, build systems from scratch that aim to introduce more features in the protocol and more flexible scripting systems that enable developers to implement features on top of the protocol \cite{Ethereum}. Summarizing, the Bitcoin technology itself or the idea underlying Bitcoin has already been successfully applied to other domains that go beyond sole payment applications. However, to the best of our knowledge, no research so far has discussed the application of Bitcoin and its underlying characteristics to the concept of S\textsuperscript{2}aaS as introduced above. 

Based on the original Bitcoin paper \cite{nakamoto2008bitcoin}, the Bitcoin-Wiki \cite{bitcoinwiki}, academic literature \cite{barber2012bitter, Meiklejohn2013} and major Bitcoin related online sources bitcoinmagazine.com, coindesk.com and thegenesisblock.com, we identified five core characteristics of the Bitcoin protocol that are relevant for S\textsuperscript{2}aaS applications: (1) decentralization and openness, (2) pseudonymous identification, (3) low fees and friction, (4) scriptability and (5) cryptographic verifiability. This process of identification and selection of characteristics was guided by our assessment of their potential impact on today's and future S\textsuperscript{2}aaS applications. A brief description of these characteristics is presented in Table \ref{table:characteristics}. These are discussed in more detail in Section \ref{sec:discussion}. We continue by introducing the basic concept of applying Bitcoin to S\textsuperscript{2}aaS with a concrete example. 

\begin{table*}[t]
\centering
\caption{Core characteristics of the Bitcoin protocol relevant for S\textsuperscript{2}aaS applications}
\begin{tabular}{l p{12cm}}
Characteristic & Description \\
\hline
\hline
Decentralization and openness & Traditional payment networks usable for S\textsuperscript{2}aaS applications always rely on a central authority and have a single point of trust. Bitcoin is an open-source decentralized network with neither a single point of trust nor failure. This may address aspects like censorship-resistance or the reduction of counter-party risks with regard to fraud but also with regard to platform independence and technology longevity.\\
\hline
Pseudonymous identification & In contrast to traditional identification or authentication systems, Bitcoin addresses are not directly connected to an identity and do not need to be registered at some central entity in the network. Unique Bitcoin addresses can be generated offline by both humans or objects and can be bound to pseudonymous or real identities whose authenticity can be proven.\\
\hline
Low fees and friction & Unlike traditional currencies and unlike traditional payment systems, Bitcoin enables transactions from anyone to anyone without the need for any intermediaries. It involves potentially neglectable transaction costs and allows for the exchange of arbitrary small units of cash in bitcoin.\\
\hline
Scriptability & Bitcoin is programmable money and as a result provides the opportunity to configure payments and data transfers directly inside of the transaction using the Bitcoin protocol. That means simple or complex conditions for cash and data transfers can be set.\\
\hline
Cryptographic verifiability & As Bitcoin is based on cryptography, methods for authentication, encryption, or integrity of the transactions are already included. As a result both humans and machines can exchange confidential data but also prove ownership with digital signatures.\\
\hline
\end{tabular}
\label{table:characteristics}
\end{table*}

\section{Example and Basic Concept}
To illustrate the here presented concept, we use a personal connected weather station like Netatmo \cite{netatmo} as an example. Personal weather stations are typically equipped with multiple sensors to continuously measure, for instance, temperature, humidity, wind speed, wind direction, solar radiation, and air pressure. Some even measure air pollution in terms of particulate matter which concerns people's health. The most obvious use for this data is to inform the owner of the weather station with the data it generates. Another example for the application of measurement data from a weather station is its use to feed the control system of the owner's household heating. For the investment into a personal weather station and heating control, its operation and maintenance, the owner gains the benefit of a well-tempered house and saves energy and money \cite{dong2014real}. 

Clearly, there is more overall use to the data, if it is shared with others. For instance, neighbours could use the exact same data to control their heating systems, however, only given the condition that the owner of the weather station is willing to share the generated data. Even for the rather simple case of a weather station, there are many other useful applications of the data. Fitness platforms like Runkeeper \cite{runkeeper.com} and Nike+ \cite{nikeplus.nike.com} could aggregate data from many owners of personal weather stations to monitor air pollution and generate running tracks with optimal air quality in real-time. Meteorologists could improve weather forecasts with the high-resolution data from many personal weather station owners. Also, researchers could use the data to track climate change \cite{JOC:JOC1276}.

There are already enthusiasts who share their data freely and without monetary incentive, e.g. \cite{wunderground.com}. However, to enable applications of greater use, there has to be extensive supply of sensor data from many weather stations all around the world. This can only be achieved by providing monetary incentive to the suppliers \cite{bohli2009initial}. In fact, monetary incentivization of sensor data providers is a main challenge faced by S\textsuperscript{2}aaS schemes to become viable in the future. As outlined above, Bitcoin technology embodies characteristics that could help to tackle those challenges. Thus far, scholars proposing S\textsuperscript{2}aaS have always introduced central coordinating intermediaries \cite{bohli2009initial} such as sensor publishers \cite{perera2014sensing} or sensor cloud platforms \cite{sheng2013sensing}. 
In the following, a completely decentralized scheme using the Bitcoin protocol is presented. First, the atomic process of exchanging a single sensor datum between two machines for cash is introduced. Limitations and extensions towards scalable implementations are discussed thereafter in Section \ref{subsec:extensions}

\subsection{Atomic Process}
\begin{figure}
\centering
\includegraphics[width=\columnwidth]{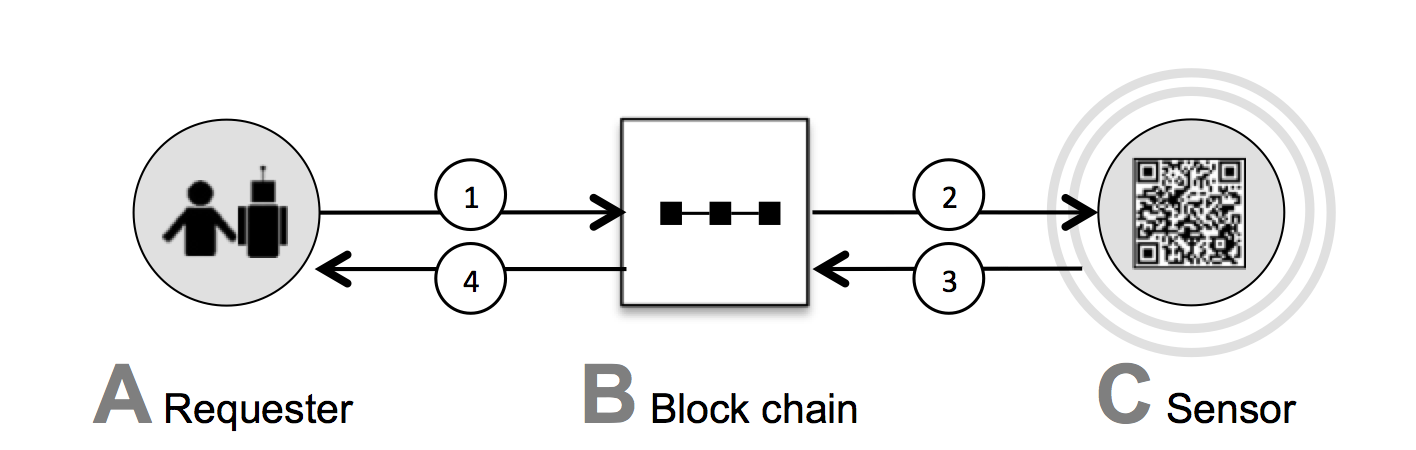}
\caption{Schema for the atomic  S\textsuperscript{2}aaS process of exchanging a single datum for cash using Bitcoin.}
\label{fig:btcDataExchange}
\end{figure}

Consider the scenario in which two machines trade a single datum for cash. The simplified process is illustrated in figure \ref{fig:btcDataExchange}, for a more in detail technical introduction please refer to \cite{bitcoinwiki, nakamoto2008bitcoin}. Given the example above the requesting machine A could be the Nike+ smartphone application which requires the current air pollution at the user's typical running track to optimize air quality during the run. The sensor C in this case would be an air pollution sensor in the vicinity of the running track. Both, the requesting machine and the sensor have a key pair which provides unique identification and allows them to transfer cash as well as private data over the block chain B as the decentralized public ledger. 
In the illustrated scenario, the requesting machine A sends a payment to the Bitcoin address (the public key\footnote{More precisely the SHA-256 hash of the public key plus a prefix.}) of sensor C. This involves the generation of a transaction that gets included in the block chain (1). In a second step, sensor C notices the receipt of the payment (2). After that, sensor C creates a transaction to the Bitcoin address of requester A, including its most current datum encrypted with A's public key  (3). Finally, requester A notices the receipt of the transaction that includes the requested datum and decrypts it using its private key (4).

\subsection{Extensions Towards Scalability}
\label{subsec:extensions}
In most cases S\textsuperscript{2}aaS applications will probably involve the transfer of more than one datum. A heating control system as well as a meteorologist needs a continuous stream of real-time data from the weather stations. Considering the instantiation of the Bitcoin protocol as it is today, the atomic scheme introduced above comes with several flaws: (1) it takes on average ten minutes until a transaction appears in the block chain, (2) every exchange involves two transactions which may entail fees, and (3) every datum has to be stored on every Bitcoin node forever. Further, there might be requests for larger datasets of historical data which cannot be included in valid transactions.

However, developments based on the inherent scriptability of Bitcoin like rapidly-adjusted (micro-) payments (see Section \ref{subsec:script}) and the possibility to transfer data off-block chain allow to build scalable solutions for S\textsuperscript{2}aaS applications. In fact, it is highly unlikely that complete sensor data will be stored in the block chain. One solution is to store it on multiple redundant third party servers and only put a pointer together with a hash of the data into the transaction. This way, the data cannot be altered later by the third party and the reference will always be in the block chain without consuming too much space.

Finally, the above scenario is based on the assumption that the requesting machine knows the sensor, the type of data it has available as well as its price. While a central repository would be a solution, it is also possible to create a decentralized sensor registry based on Namecoin much like OneName is a decentralized registry for people \cite{onename.io}.

\section{Discussion}
\label{sec:discussion}

In this section, we revisit the S\textsuperscript{2}aaS relevant characteristics of the Bitcoin protocol introduced in table \ref{table:characteristics} and discuss their application to S\textsuperscript{2}aaS schemes. 

The core Bitcoin protocol is continuously improved and extended by a vibrant open source community. Developers are actively adapting the core protocol to the needs of the rapidly growing eco-system. As the Internet was not originally designed for the applications that we use today, it still evolved on top of the original underlying protocols like TCP/IP and HTTP. For this reason we choose not to focus on specific single technical limitations of the current Bitcoin protocol implementation, but rather argue from a broader technological perspective.

\subsection{Decentralization and Openness}

Comparable to HTTP as a protocol for transfer of data, Bitcoin is a protocol for the transfer of cash on the Internet. It is based on a decentral design and consequently has no single point of trust. Just like HTTP, anyone is free to use it and build applications on top of it. This has important ramifications for its use in S\textsuperscript{2}aaS applications.

As Bitcoin is a peer-to-peer network, it is censorship-resistant. Consequently, no central authority can systematically exclude someone or something from participating. This represents a crucial difference to classical payment networks (e.g. Visa, MasterCard, or PayPal) that can ban anyone from using their services (as happened to Wikileaks in 2010 and Russian bank customers in 2014). Using Bitcoin as payment layer for S\textsuperscript{2}aaS application brings censorship-resistance to sharing sensor data for cash payment. Nobody could systematically be excluded to buy or sell sensor data.

Without counter-party risk of an intermediary to process both payments and data transfer, applications leveraging on a Bitcoin enabled S\textsuperscript{2}aaS  environment do not carry the risk of self-interested (even justified) policy changes by central entities. For instance, policy changes by Twitter forced some third-party developers using the Twitter API to shut down their operations \cite{twitterAPI}. Something like this cannot happen using Bitcoin, as there is no central authority able to change the rules out of self-interest. Using Bitcoin as a payment network is completely platform independent. This should give entrepreneurs, sensor data providers (like the personal weather station owners in our example) and established platforms alike confidence in the stability, longevity and availability of S\textsuperscript{2}aaS services built on top of the Bitcoin protocol and stir innovation.

Further, Bitcoin is an open source project with many readily available implementations that can be adapted and extended (see bitcoin.org/en/development for a list). Anyone is free to use Bitcoin technology and innovate with it. This gives Bitcoin a head start for innovation which will continue to grow rapidly all areas that require some sort of financial service, like payment, including S\textsuperscript{2}aaS applications.

\subsection{Pseudonymous Identification}

A viable S\textsuperscript{2}aaS network requires that all entities have  to be  uniquely  identified  and  authenticated. Bitcoin addresses can be assigned to all types of entities in a S\textsuperscript{2}aaS network. Ownership of an address is provable with public key cryptography and allows for pseudonymous authentication.

As Bitcoin addresses are not directly connected to an identity and do not need to be registered at some central entity in the network, they guarantee pseudonymity of the owner. This can be favorable for S\textsuperscript{2}aaS applications, because data providers may not want to expose their identity. However, it does not necessarily mean that owners are anonymous. As all transactions are publicly available on the block chain, any payment can be traced to an address that can possibly be connected to an identity at some point.

Bitcoin addresses cannot only be assigned to persons and used as an equivalent to a bank account. Objects like cars, fridges, houses and - like in our example -- personal weather stations -- can have a Bitcoin address, effectively enabling them to send and receive cash. As the Bitcoin network does not incorporate any intermediaries, Bitcoin transactions can be carried out completely automated. In fact, Bitcoin transactions can be carried out equally well by machines and humans enabling direct machine-to-machine payments. We believe, this will have strong impact on future S\textsuperscript{2}aaS applications and the IOT in general.

\subsection{Low Fees and Friction}

Bitcoin is a global payment network enabling cash transactions from anyone (or anything) to anyone (or anything) at any time directly without the need for an intermediary. Thereby, Bitcoin can scale down payments to very low amounts allowing for trade of very small exchangeable units. We believe, low fees and friction in the Bitcoin payment network can lead to a whole wave of IOT innovations, because the programmatic exchange of arbitrary amounts of cash without human intervention and intermediaries allows for a generation of IOT applications that has not been feasible before.

Using Bitcoin technology as a payment network for S\textsuperscript{2}aaS applications may allow for purchases of single data points, as in our basic example of a personal weather station, costing way less than the smallest available units of any traditional currency. This would not be possible using traditional payment network, at least not without the introduction of additional processes. Typical intermediaries in a classical payment network like Visa, MasterCard collect fees of one to three percent \cite{chakravorti2003theory}. Average fees for international remittances (e.g. with Western Union or MoneyGram) even account for more than eight percent of the transaction amount \cite{remittances}.

Bitcoin transactions are not free, but rather compete against each other for space in the block chain in something similar to a bidding process. Senders of Bitcoin transactions can include a voluntary, so called miner fee with their transactions. Transactions with higher fees are given higher priority by miners and are consequently processed faster than transactions with lower fees (This is a simplified description. For an accurate description of the transaction prioritization mechanism, please refer to \cite{bitcoinwiki, nakamoto2008bitcoin}). By exposing Bitcoin transactions to these simple supply and demand market dynamics transaction costs are no longer dictated by the gatekeepers of payment networks and price efficiency in the processing of transactions for all types of applications will eventually be established.

\subsection{Scriptability}
\label{subsec:script}

Bitcoin is programmable money in the sense that transactions are scriptable. For instance, the validity of a transaction message, and hence its clearing, can be bound to certain conditions. By combining multiple transaction messages and conditions, rather complex contracts requiring no trust between the parties can be established. Traditionally such contracts are enforced by intermediaries. Bitcoin allows to establish contracts and enforce them completely without intermediaries in the block chain. The following examples are taken from \cite{smartcontr}. We briefly describe the different types of scripted transactions and put them into context to S\textsuperscript{2}aaS.

\subsubsection{Multi-signature Transactions}
Bitcoin allows to require several parties to sign a transaction in order to claim an output. This type of transaction is the basis for many use cases as can be seen in the following.
\subsubsection{Escrow and Dispute Mediation}
In the case a requester wants to buy data from an untrusted sensor, a third party can act as an escrow service by using multi-signature transactions. Notably, the mediator can never claim the bitcoins by itself. This can be useful to hold back payments until the received data can be validated, like in cases where the value of data is closely linked to additional actions or events, and needs to go through some kind of validation phase. 
\subsubsection{Assurance Contracts}
Assurance contracts can be used to fund public goods in a similar fashion to a crowd-funding campaigns (cf. Kickstarter.com or Indiegogo.com) but without the need for a third party. For instance, an entrepreneur could offer to build a sensor network to measure air quality in a city if s/he is able to raise a particular amount of money. People could then send transactions to the entrepreneur privately with their pledge as inputs and the output to be the fund limit. Therefore, the transactions become only valid if the entrepreneur collects enough transactions to meet the funding. In the S\textsuperscript{2}aaS case, assurance contracts may play a role to crowd-fund data that is useful to many, but could easily be duplicated and given away for free. The provider could couple the release of data to a sufficient payment threshold.
\subsubsection{Using External State}
With multi-signature transactions an oracle can be required to sign transactions. The oracle is a server with a key pair that only signs a transaction if a particular expression returns true. The incorporation of external state in Bitcoin transactions could lead to the development of decentralized trustless prediction markets, requiring a S\textsuperscript{2}aaS network as oracle. For instance, movie producers could hedge against bad weather on a filming day. The external state which resolves the bet in the block chain could stem from the multiple personal weather stations in our example.
\subsubsection{Rapidly-Adjusted (Micro-)Payments}
Although Bitcoin allows to transfer tiny fractions of a bitcoin with little or no fees, several problems arise if this is done in fast succession. However, using time lock and multi-signature transactions it is possible to create micropayment channels where payments between two parties are aggregated without the need for a trusted third party. In the S\textsuperscript{2}aaS scenario such a payment channel can be used if a requester wants to subscribe to a sensor for a specific amount of time.

\subsection{Cryptographic Verifiability}
Bitcoin uses digital signatures to prove ownership of addresses. Using the same technique, persons, or in our case sensors, can authenticate themselves by signing a message, proving ownership of their Bitcoin address. Confidentiality of the data to be sent can be guaranteed by encrypting the data with the public key of the receiver. Further, the block chain can be used as a notary service for the transacted data. Hashes of the data can be stored in the block chain to prove that the data was not altered later on.

\section{Conclusion and Outlook}

The IOT is expected to consist of billions of sensor nodes bridging the gap between the physical and digital world. Based on the idea that not only the one who generates data can profit from it, the concept of S\textsuperscript{2}aaS foresees global sensor data markets. However, to carry this idea from theory to practice, there are manifold systemic hurdles to overcome. Naturally, a low cost micropayment system for sensor data has to be in place, so sellers of data are able to receive monetary gratification from the buyer side. Sensors and its owners have to be uniquely identified and authenticated and values delivered by the sensors should be traceable and have to be secured against manipulation. 

In this paper, we introduced Bitcoin with its core characteristics and discussed how IOT applications and more specifically S\textsuperscript{2}aaS applications could benefit from Bitcoin technology or its conceptual idea. We derived five core characteristics that could drive innovation in this context: (1) decentralization and openness, (2) pseudonymous identification, (3) low fees and friction, (4) scriptability, and (5) cryptographic verifiability. We discuss how the integration of these characteristics in S\textsuperscript{2}aaS applications may trigger the right business dynamics that are needed on the buyer side and on the seller side of emergent sensor data by creating value for all exchange partners that is solely based on fair market dynamics defined in the underlying protocol. 

As the eco-system around Bitcoin is growing and its technology and conceptual idea has already been transferred successfully to other domains and purposes, it is our belief that it could also drive the development of S\textsuperscript{2}aaS business models. These could build on the innovations that the integration of discussed characteristics provides and offer new value propositions. It is our intent to start the conversation with IOT scholars and practitioners. From our perspective, future research should focus on bringing Bitcoin and the IOT closer together in two ways. First, by further refining the core characteristics of Bitcoin in the IOT context and eventually building a framework for IOT applications to further understand Bitcoin's potential role in its evolution. Second, by conceptualizing IOT applications using Bitcoin, solving known problems and inventing new applications.

\bibliographystyle{IEEEtran}
\bibliography{lit}

\end{document}